\documentclass[a4paper]{jpconf}
\usepackage{graphicx}
\usepackage{lineno}
\usepackage{cite}
\usepackage{floatrow}
\makeatletter \providecommand{\citedash}{\hbox{$\sim$}\penalty\@m}
\makeatother
\begin{document}
\title{Higher Moments of Net-Kaon Multiplicity Distributions at STAR}

\author{Ji Xu (for the STAR Collaboration)}

\address{ Key Laboratory of Quark \& Lepton Physics (MOE) and Institute of Particle Physics, Central China Normal University, Wuhan 430079, China

   Lawrence Berkeley National Laboratory, Berkeley, CA 94720, US}

\ead{jixu@lbl.gov}

\begin{abstract}
Fluctuations of conserved quantities such as baryon number (B), electric charge number (Q), and strangeness number (S), are sensitive to the correlation length and can be used to probe non-gaussian fluctuations near the critical point. Experimentally, higher moments of the multiplicity distributions have been used to search for the QCD critical point in heavy-ion collisions. In this paper, we report the efficiency-corrected cumulants and their ratios of mid-rapidity ($|y|<0.5$) net-kaon multiplicity distributions in Au+Au collisions at $\sqrt{s_{ \rm NN}}$ = 7.7, 11.5, 14.5, 19.6, 27, 39, 62.4, and 200 GeV collected in 2010, 2011, and 2014 with STAR at RHIC. The centrality and energy dependence of the cumulants and their ratios, are presented. Furthermore, the comparisons with baseline calculations (Poisson) and non-critical-point models (UrQMD) are also discussed.
\end{abstract}

\section{Introduction}The main goals of the high energy nuclear collisions are to explore the phase structure of strongly interacting hot and dense nuclear matter and map the quantum chromodynamics (QCD) phase diagram which can be displayed in the two dimensions (temperature (T) vs. baryon chemical potential ($\mu_{B}$)).
Fluctuations of conserved quantities, such as net-baryon ($\Delta{N_{B}}$), net-charge ($\Delta{N_{Q}}$) and net-strangeness ($\Delta{N_{S}}$),  have been predicted to be sensitive to the QCD phase transition and the QCD critical point~\cite{SearchCP1,SearchCP2,SearchCP4}. The cumulants of the event-by-event distributions of these conserved quantities are directly connected to the thermodynamic susceptibilities computed with Lattice QCD~\cite{LQCD1,LQCD2,LQCD3} and in the Hadron Resonance Gas (HRG) model~\cite{HRG1,HRG2,HRG3,HRG4}. Thus, those cumulant ratios are equivalent to the ratios of various order susceptibilities as $C_{4}/C_{2} = \chi_{i}^{(4)}/\chi_{i}^{(2)}$ and $C_{3}/C_{2} = \chi_{i}^{(3)}/\chi_{i}^{(2)}$, where $i$ indicates the conserved quantity. 

 
Experimentally, we use net-kaon ($\Delta{N_{K}}$) as proxy for net-strangeness.
The fluctuations of conserved quantities have been widely studied experimentally and theoretically~\cite{HRG1,SearchCP1,SearchCP2,SearchCP3,SearchCP4,SearchCP5,SearchCP6,xflQM2015}. 
In this work, we report recent efficiency-corrected cumulants and cumulant ratios of the net-kaon ($\Delta{N_{K}}$) multiplicity distributions measured in Au+Au collisions at $\sqrt{s_{\rm NN}}$ = 7.7, 11.5, 14.5, 19.6, 27, 39, 62.4, and 200 GeV collected in 2010, 2011, and 2014 by STAR at RHIC.

\section{Analysis Details}
The STAR (Solenoidal Tracker At RHIC) detector at BNL has a large uniform acceptance at mid-rapidity and excellent particle identification capabilities.
Energy loss ($dE/dx$) in the Time Projection Chamber and mass-squared ($m^{2}$) from the Time-Of-Flight detector are used to identify kaons.
The $K^{+}$ ($K^{-}$) for net-kaon analysis are measured at mid-rapidity ($|y|$$<$0.5) within the transverse momentum region 0.2$<$$p_{T}$$<$1.6 GeV/c.

For the centrality selection, the uncorrected multiplicity distribution of primary charged particles, excluding kaons/anti-kaons within $|\eta|$$<$1.0 has been used for $\Delta{N_{K}}$ to avoid auto-correlations.
The results are corrected for a finite centrality bin width~\cite{technique}.

\begin{figure*}
\hspace{0cm}
\caption{Uncorrected raw event-by-event net-kaon ($\Delta{N_{K}}$) multiplicity distributions for Au+Au collisions at $\sqrt{s_{\rm NN}}$ = 7.7, 11.5, 14.5, 19.6, 27, 39, 62.4, and 200GeV for 0-5\% top central (black circles), 30-40\% central (red squares), and 70-80\% peripheral collisions (blue stars).}
\centerline{\includegraphics[width=4in]{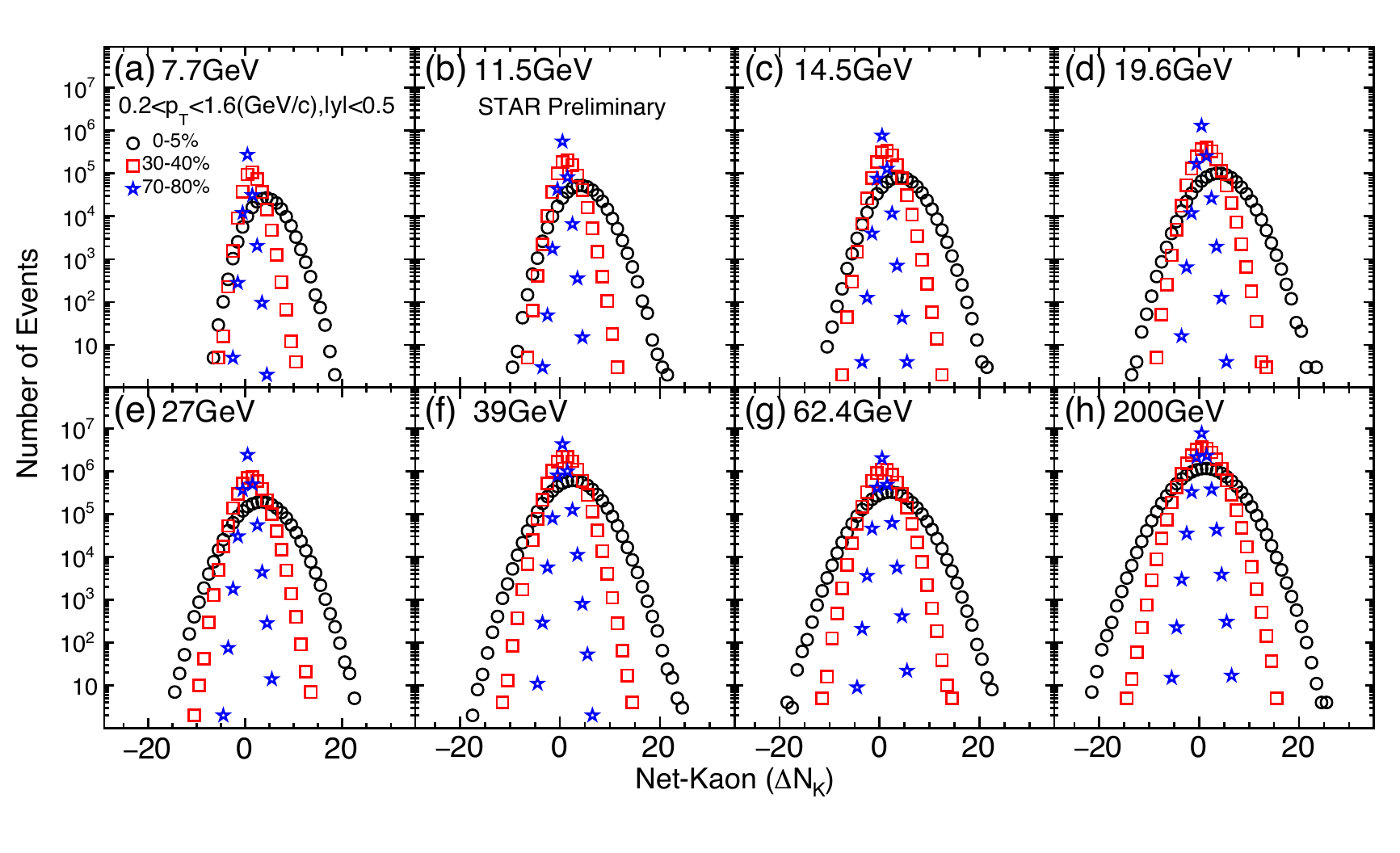}}
\label{distributions}
\end{figure*}

Figure~\ref{distributions} shows the uncorrected event-by-event net-kaon multiplicity distributions for Au+Au collisions at $\sqrt{s_{\rm NN}}$ = 7.7, 11.5, 14.5, 19.6, 27, 39, 62.4, and 200 GeV for $\Delta{N_{K}}$ in three centrality intervals. We observe that most central collisions have a wider distribution compared with peripheral collision. The peak of the net-kaon distributions shift slightly towards to the positive direction with the energy decrease. Detailed discussions about the efficiency correction and error estimation can be found in~\cite{technique,error}.

\section{Results and Discussion} 

\begin{figure*}
\centerline{\includegraphics[width=5in]{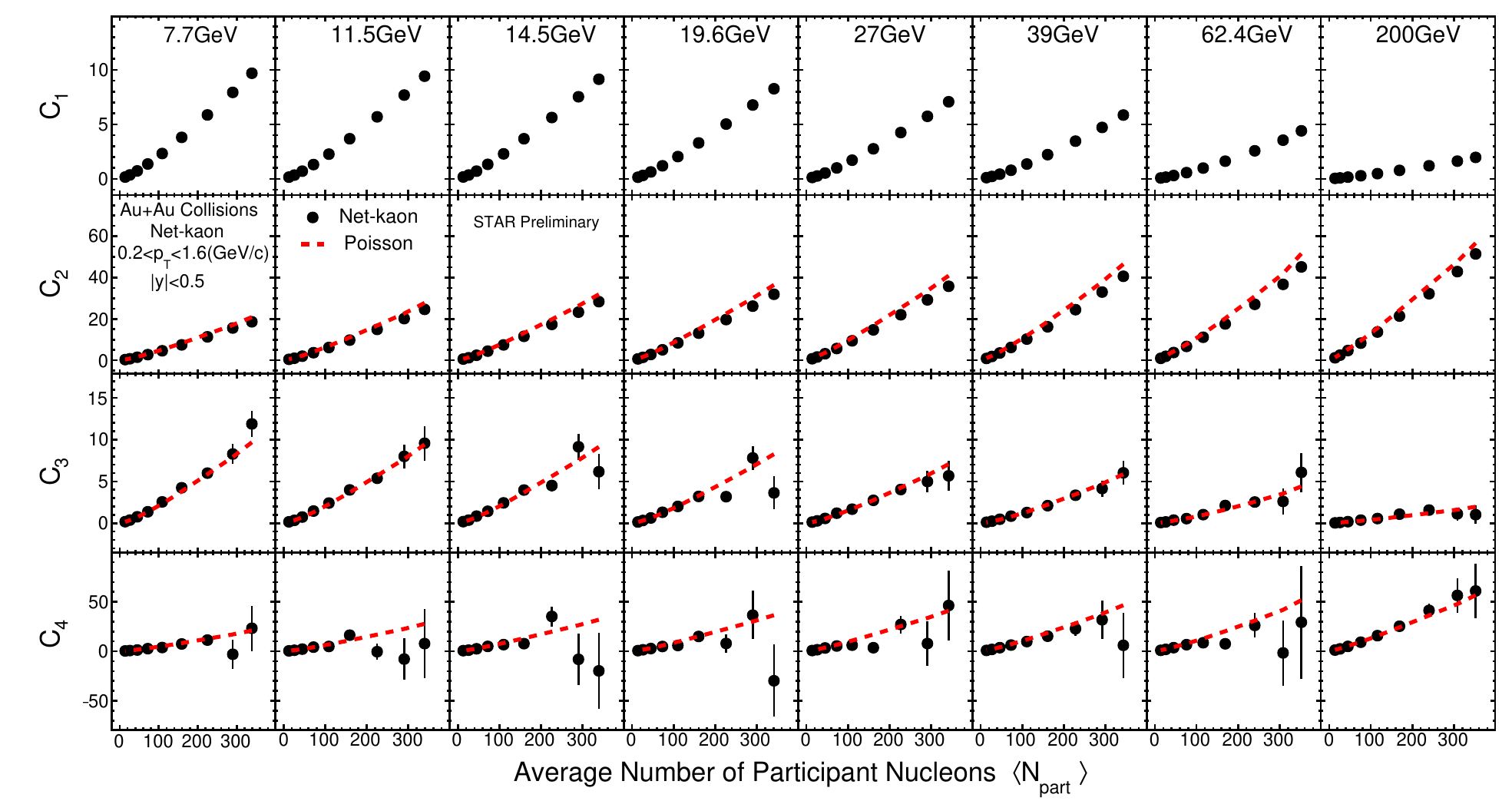}}
\caption{Centrality dependence of cumulants $(C_{1}, C_{2}, C_{3}$, and $C_{4})$ of net-kaon multiplicity distributions for Au+Au collisions at $\sqrt{s_{\rm NN}}$ = 7.7, 11.5, 14.5, 19.6, 27, 39, 62.4, and 200GeV. The Poisson expectations are denoted as dotted lines. The error bars are statistical errors.}
\label{cumu_egy}
\end{figure*}

\begin{figure*}
\centerline{\includegraphics[width=3.2in]{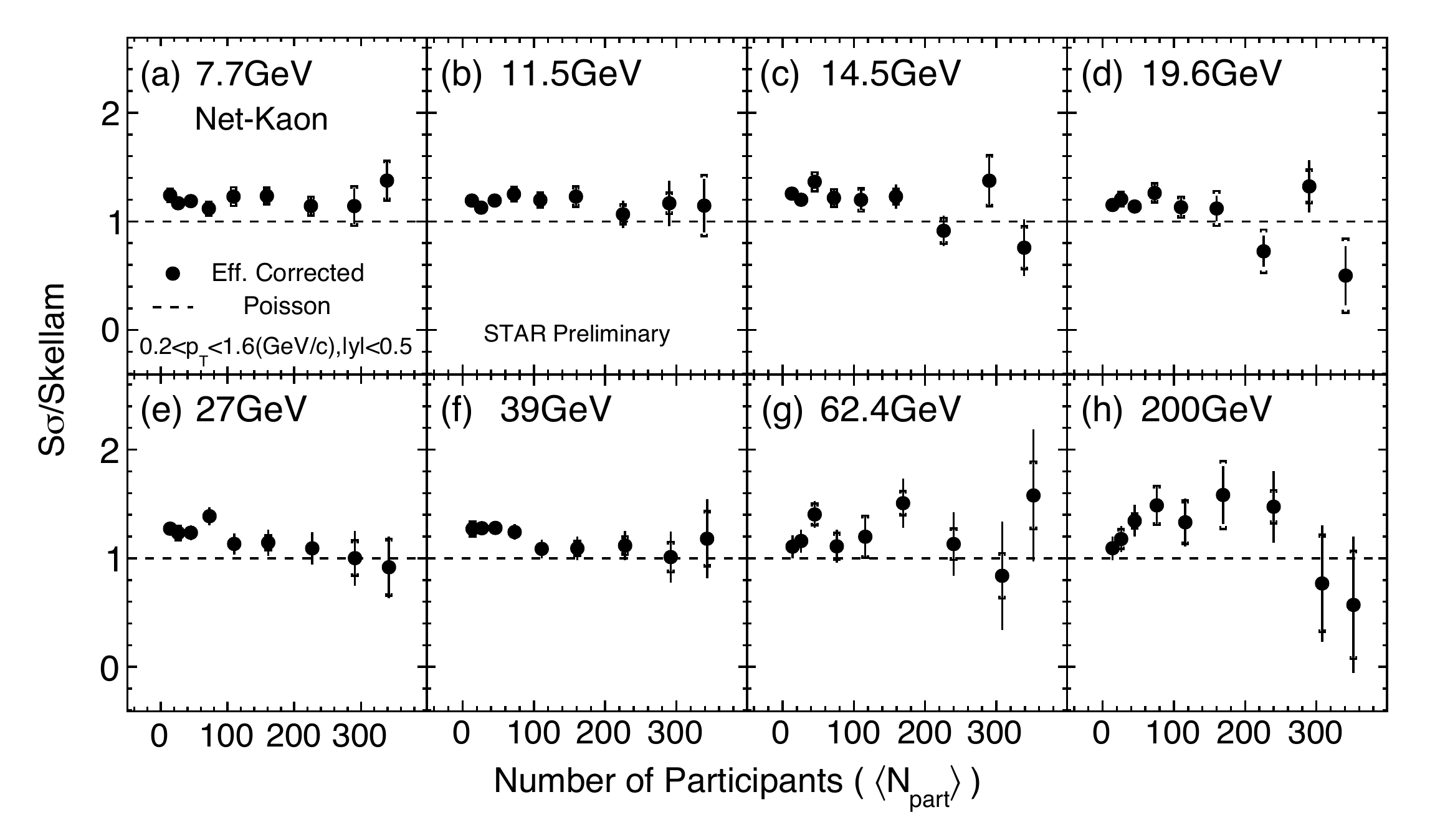}
\includegraphics[width=3.2in]{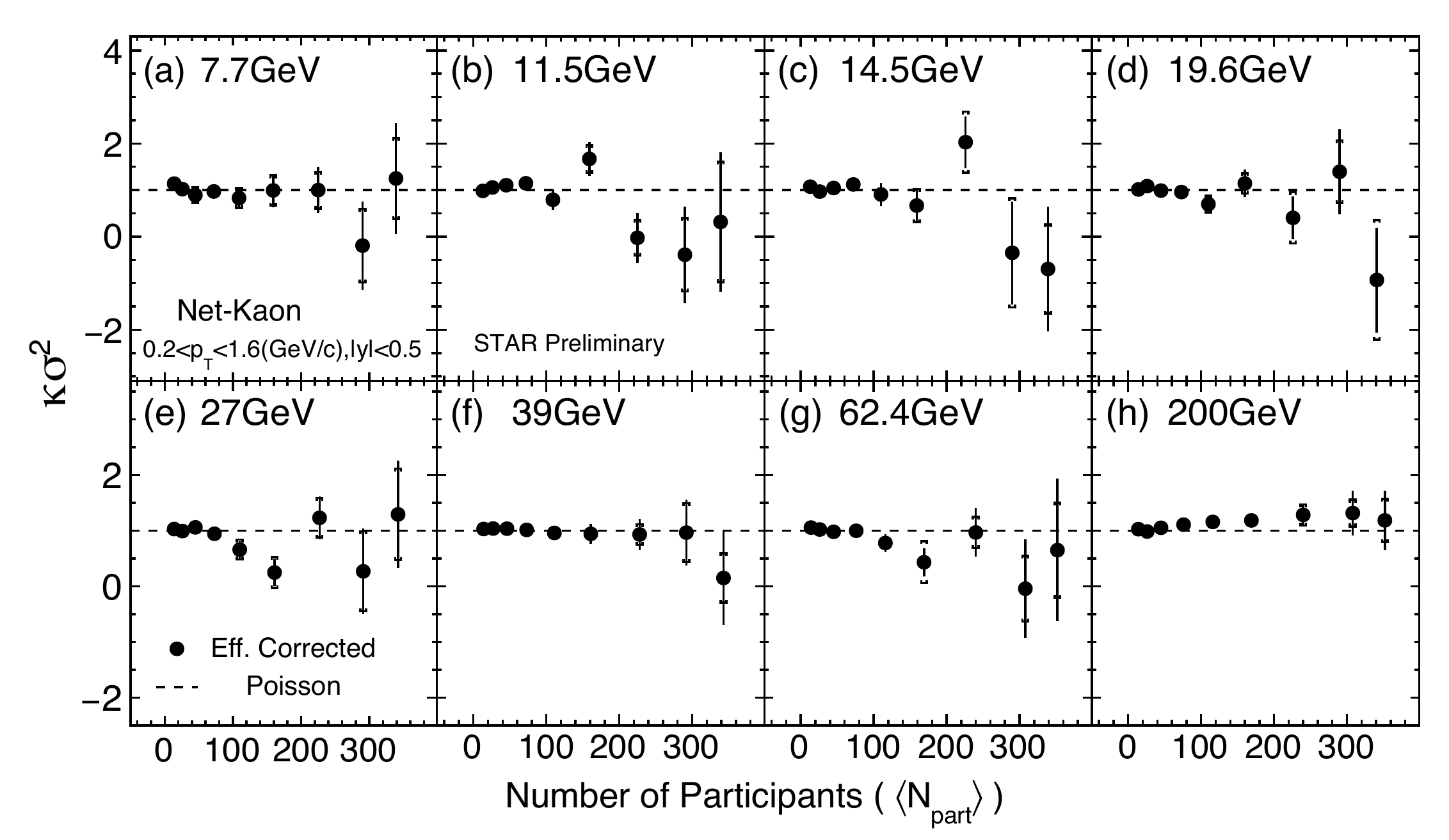}}
\caption{ Centrality dependence of $S\sigma/$Skellam (Left) and $\kappa\sigma_{2}$ (Right) of net-kaon multiplicity distributions for Au+Au collisions at $\sqrt{s_{\rm NN}}$ = 7.7, 11.5, 14.5, 19.6, 27, 39, 62.4 and 200GeV. The Poisson expectations are denoted as dotted lines. The error bars are statistical errors and the caps represent systematic errors.} 
\label{SDKV}
\end{figure*}

\begin{figure*}
{\includegraphics[width=2.3in]{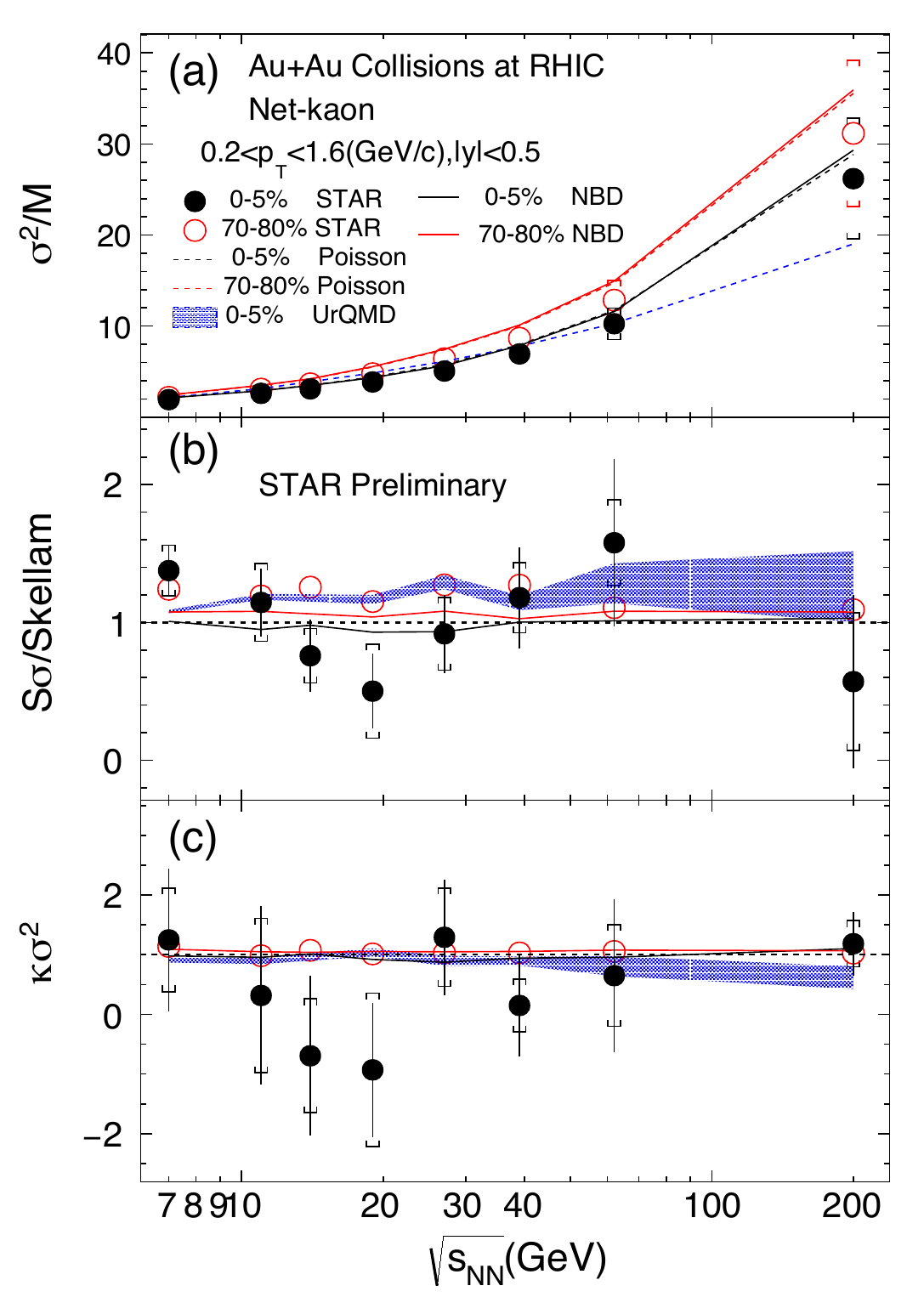}}
\caption{Energy dependence of cumulant ratios ($\sigma^{2}$/M, $S\sigma$/Skellam, $\kappa\sigma^{2}$) of net-kaon multiplicity distributions for top 0-5\% central and 70-80\% peripheral collisions at Au+Au collisions at $\sqrt{s_{\rm NN}}$ = 7.7, 11.5, 14.5, 19.6, 27, 39, 62.4, and 200GeV. 
The Poisson and NBD expectations are denoted as dotted and solid lines, UrQMD calculations are shown as blue bands. The error bars are statistical errors and the caps represent systematic errors.
}
\label{energy}
\end{figure*}

All results have been corrected for finite detector reconstruction efficiencies, with the statistical uncertainties calculated based on the Delta Theorem~\cite{error}.
Figure~\ref{cumu_egy} shows the centrality dependence of cumulants ($C_{1}$, $C_{2}$, $C_{3}$, and $C_{4}$) of net-kaon multiplicity distributions in Au+Au collisions at $\sqrt{s_{\rm NN}}$ =7.7$\sim$200 GeV. The red dashed lines represent the Poisson expectation. In general, those cumulants show a linear variation with the averaged number of participant nucleons and most $C_{3}$ and $C_{4}$ values are consistent with Poisson expectation within uncertainties. 

Figure~\ref{SDKV} shows the centrality dependence of $S\sigma/$Skellam and $\kappa\sigma_{2}$ of net-kaon multiplicity distributions for Au+Au collisions at $\sqrt{s_{\rm NN}}$ = 7.7 to 200GeV. The error bars are statistical errors and the caps represent systematic errors. For $S\sigma/$Skellam, the values are consistent with Poisson expectation for most central collisions within uncertainties. Most of the values of $\kappa\sigma^2$ are consistent with Poisson baseline, which is unity, within uncertainties.

The energy dependence of the volume-independent cumulant ratios ($\sigma^{2}/M$, $S\sigma/$Skellam, $\kappa\sigma^2$) for net-kaon multiplicity distributions in Au+Au collisions are presented in Figure~\ref{energy}. The blue bands give the predictions from the UrQMD calculations which do not include critical physics processes.
One can see that the values of $\sigma^{2}$/M increase as the energy increases. The values of $S\sigma$/Skellam and $\kappa\sigma^{2}$ for net-kaon multiplicity distributions are consistent with both poisson and negative binomial distribution baseline within errors. Moments results from UrQMD (no Critical Point) show no energy dependence for $S\sigma$/Skellam and $\kappa\sigma^{2}$.

The measurements presented here will be further improved in the upcoming RHIC BES II in 2019-2020 which include an STAR detector upgrade. An inner TPC and Endcap TOF upgrade will enlarge the phase-space up to $|\eta|$$<$1.5 and down to $p_{T}$ = 60 MeV/c. The Event Plane Detector at forward rapidities will allow for a better centrality estimation, suppressing auto-correlations. 

\section{Summary}
In this paper, centrality and energy dependence of cumulants (up to forth order) and their ratios from $\sqrt{s_{\rm NN}}$ = 7.7 to 200 GeV in Au+Au collisions for net-kaon multiplicity distributions are presented. Within uncertainties (statistical and systematic),  $S\sigma$/Skellam and $\kappa\sigma^{2}$ values are consistent with the Poisson and negative binomial distribution baseline within errors at most central collisions. The results from UrQMD model calculations, where no-criticality is intended, show no energy dependence.
The RHIC BES II will bring a larger event sample and a wider phase-space to enhance the search for the QCD critical point in the high net-baryon density region 420 $< \mu_{B} <$ 250 MeV (7.7 $< \sqrt{s_{\rm NN}} <$ 19.6 GeV) .

\section*{Acknowledgments}
The work was supported in part by the MoST of China 973-Project No.2015CB856901, NSFC
under grant No. 11575069, 11221504.

\section*{References}
\bibliographystyle{apsver4-1}

\end{document}